\documentclass{article}
\usepackage{spconf,amsmath,graphicx,hyperref}
\usepackage{amsfonts}
\usepackage{booktabs,multirow,array}
\usepackage{etoolbox}

\renewenvironment{thebibliography}[1]{%
  \begin{oldthebibliography}{#1}%
  \renewcommand{\baselinestretch}{1}%
  \fontsize{9}{10}\selectfont
  \setlength{\itemsep}{0pt}%
  \setlength{\parsep}{0pt}%
  \setlength{\parskip}{0pt}%
}{%
  \end{oldthebibliography}%
}

\title{TF-MoE-USS: Time-Frequency Mixture-of-Experts \\
for Universal Sound Separation}
\name{Bohan Liu$^{1,*}$, Tianhua Li$^{1,*}$, Chenda Li$^{1,2}$, Wangyou Zhang$^{1,2}$, Yanmin Qian$^{1,2}$\thanks{
OpenAI Codex assisted with figure and table preparation and language polishing. The authors reviewed and verified all content and take full responsibility for the final manuscript.
  $^*$ represents equal contribution.}}

\address{$^1$Auditory Cognition and Computational Acoustics Lab, \\ Shanghai Jiao Tong University, Shanghai, China \\
$^2$VUI Labs \\
}
\begin{document}
\ninept
\maketitle
\begin{abstract}
The acoustic diversity of universal sound separation (USS) calls for models that balance capacity and computational cost. We propose TF-MoE-USS with two complementary contributions. First, we extend TF-MoE from speech separation to USS, using its load-balanced time- and frequency-axis routing to expand model capacity with limited additional computation. Second, we introduce event-supervised mixture-level routing, which uses source-event labels to guide expert selection during training and complements acoustic routing with mixture-level semantic information. 
Both routing mechanisms use top-1 expert selection and require no labels or prompts at inference.
Trained on 2,000 hours of source audio, TF-MoE-USS achieves 8.42 dB signal-to-distortion ratio (SDR), surpassing state-of-the-art models with 84.8\% fewer multiply-accumulate operations.~\footnote{Demo: \url{https://audiocc-lab.github.io/tfmoe-uss/}}

\end{abstract}


\section{Introduction}
\label{sec:introduction}


Universal sound separation (USS) recovers individual sources from arbitrary sound mixtures, extending beyond speech or music to diverse acoustic events. Its central challenge is capturing widely varying spectral structures and temporal dynamics across sound classes within a shared model~\cite{kavalerov2019universal}. Transient events, harmonic sounds, and noise-like textures can overlap in time and frequency, so separation cues vary across sources within the same mixture. Progress has been supported by diverse training and evaluation data~\cite{wisdom2021fuss,li2026hive} and large-scale training across speech, music, and sound events~\cite{pons2024gass}. Beyond supervised objective refinement~\cite{postolache2023adversarial}, Mixture Invariant Training (MixIT)~\cite{wisdom2020mixit} and weakly labelled learning~\cite{chen2022zeroshotuss,kong2026universalweak} broaden the use of real mixtures and coarsely annotated recordings for training. Self-supervised audio pretraining also provides representations that complement spectral features for separation~\cite{zhao2024amaeuss}. Handling this diversity requires sufficient model capacity to represent both shared and class-dependent acoustic structure.


Many USS systems build on architectures originally developed for speech and music separation, where discriminative dense separators directly estimate sources from mixtures~\cite{luo2023bsrnn,wang2023tfgridnet}. Although efficient architectures reduce the cost of unified separation~\cite{paissan2025fastuss}, increasing dense model capacity generally increases active computation, making larger models costly to train and deploy. Query-based extraction further provides explicit source selection through language or multimodal cues~\cite{li2023crossmodality}. For example, CLAPSep~\cite{ma2024clapsep} and AudioSep~\cite{liu2025audiosep} combine query representations with separation features through FiLM~\cite{perez2018film}. This flexibility introduces query-encoding and conditioning components, while single-target systems may require repeated separation passes to recover multiple sources. Joint prediction can recover several specified sources in one pass~\cite{saijo2025tuss}, but remains conditioned on source prompts. Iterative generative extraction adds further network evaluations and latency~\cite{yuan2025flowsep,shi2026samaudio}. These limitations motivate end-to-end multi-source USS that jointly recovers sources from the mixture in a single forward pass, without external prompts or auxiliary models at inference, while scaling capacity without proportional computational growth.

\begin{figure*}[!ht]
    \centering
    \includegraphics[width=\textwidth]{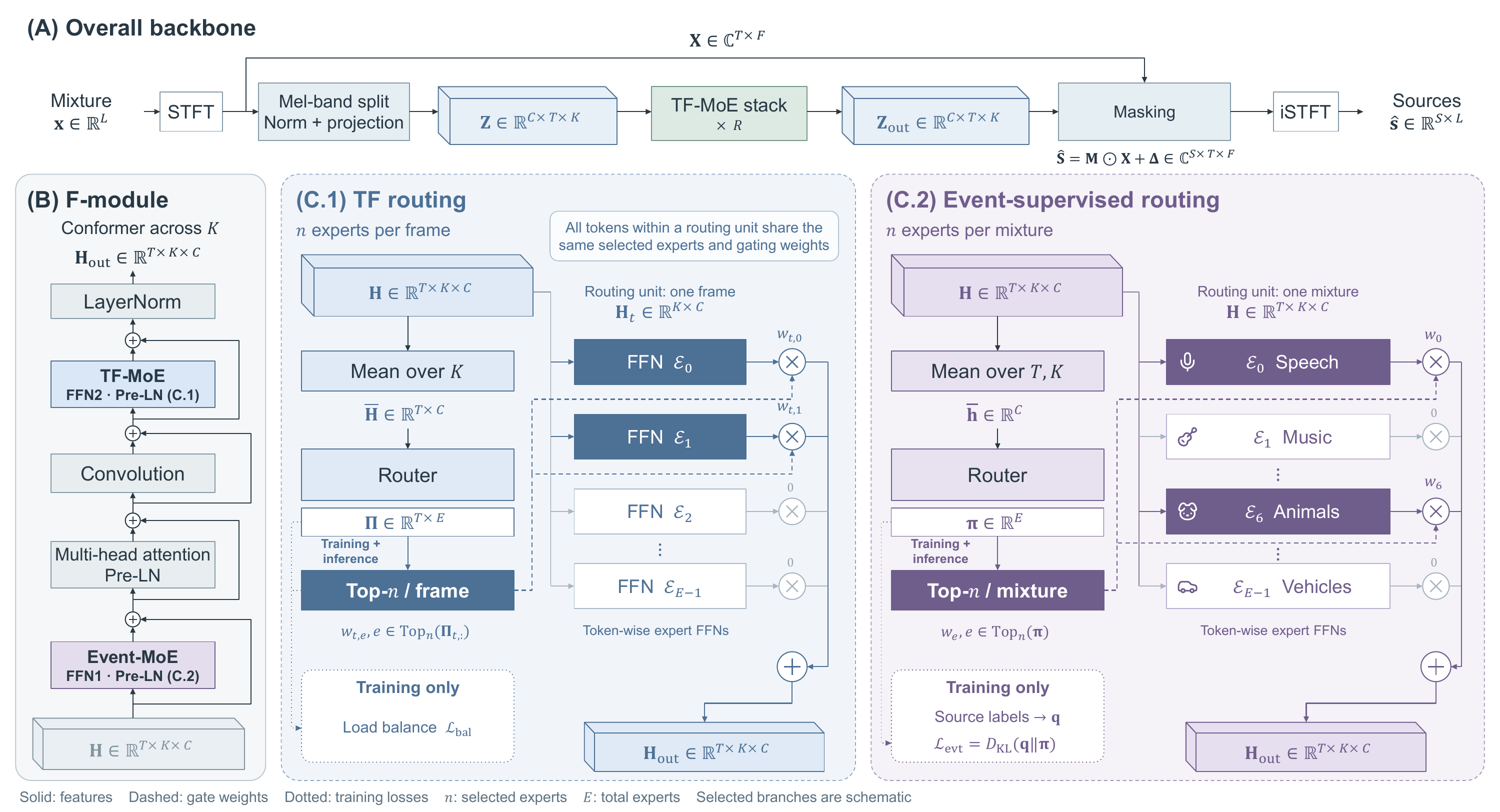}
    \caption{Overview of our method. (A) Overall architecture. (B) Internal architecture of the F-module Conformer block. (C.1) MoE routing mechanism under TF-level routing. (C.2) MoE routing mechanism under event-level routing.}
    \label{fig:pipeline}
\end{figure*}

We address this need using sparse mixture of experts (MoE)~\cite{shazeer2017moe,fedus2022switch}, which expands model capacity while activating only a few experts, limiting additional computation. TF-MoE~\cite{hu2026tfmoe} applies this mechanism to speech separation, but its acoustic routing lacks explicit supervision for sound-event specialization. Semantic information has been explored through joint separation and classification~\cite{kong2018jointseparation,tzinis2020classification,lee2025deftmamba} and auxiliary losses in MixIT~\cite{wisdom2021sparsemixit}. Multiple-Embedding Separation Networks~\cite{munakata2021multiple} learn sound-class-specific features but evaluate all embedding networks during inference. We introduce event-supervised routing that applies source-event supervision directly to the sparse routing distribution, guiding expert allocation with mixture-level semantics that complement local time-frequency cues. The event router pools features over time and frequency to obtain mixture-level context. During training, source-event labels are mapped to one-hot expert assignments and averaged across sources into a permutation-invariant soft routing target. At inference, top-1 expert selection uses only mixture features, without event labels or an auxiliary semantic model. Experiments demonstrate improved separation at comparable computation and substantially lower inference computation at comparable performance. Our main contributions are as follows:
\begin{itemize}
    \item We extend TF-MoE from speech separation to USS, expanding model capacity through sparse time- and frequency-axis routing with limited additional computation.
    \item We introduce event-supervised mixture-level routing to align expert allocation with sound events, complementing acoustic routing without inference-time labels or prompts.
    \item We demonstrate improved separation across data and model scales. Our best model achieves 8.42 dB SDR on the three-source Hive test set with 84.8\% fewer multiply-accumulate operations than state-of-the-art baselines.
\end{itemize}


\section{Method}
\label{sec:method}

\subsection{TF-MoE Backbone}
\label{ssec:backbone}

We build on TF-MoE's mel-band separator~\cite{hu2026tfmoe} and Conformer backbone~\cite{gulati2020conformer}, adding sound-event supervision for routing. In Fig.~\ref{fig:pipeline}(A), the short-time Fourier transform (STFT) maps mixture waveforms $\mathbf{x}\in\mathbb{R}^{L}$ to complex spectra $\mathbf{X}\in\mathbb{C}^{T\times F}$. Here, $L,T,F$ denote the waveform length, frame count, and frequency-bin count, respectively. Mel-band splitting, normalization, and projection yield $\mathbf{Z}\in\mathbb{R}^{C\times T\times K}$, with $C$ channels and $K$ bands.

Each of the $R$ TF-MoE blocks applies a frequency Conformer (F-Module) across $K$, then a temporal Conformer (T-Module) across $T$, preserving the shape in $\mathbf{Z}_{out}$.
Following~\cite{High-Yu2023}, the masking module produces
complex mask and residual $\mathbf{M},\boldsymbol{\Delta}\in\mathbb{C}^{S\times T\times F}$ for $S$ sources:
\begin{equation}
 \hat{\mathbf{S}}=\mathbf{M}\odot\mathbf{X}+\boldsymbol{\Delta}.
 \label{eq:reconstruction}
\end{equation}
The product is complex elementwise multiplication, with source-axis broadcasting. Inverse STFT converts $\hat{\mathbf{S}}$ into waveforms $\hat{\mathbf{s}}\in\mathbb{R}^{S\times L}$. Each Conformer module contains attention and convolution between two pre-normalized feed-forward networks (FFNs), with $1/2$-scaled FFN residuals and final layer normalization.

\subsection{Axis Routing}
\label{ssec:axis_training}

Fig.~\ref{fig:pipeline}(B) shows the F-module with pre-normalized FFN input $\mathbf{H}\in\mathbb{R}^{T\times K\times C}$; the T-module applies the same routing and Conformer operations with $T,K$ exchanged, then restores their order. As shown in Fig .\ref{fig:pipeline}(C.1), each routed FFN has $E$ independent experts $\mathcal{E}_e$, $e\in\{0,\ldots,E-1\}$, comprising Linear ($C\to4C$), Swish~\cite{ramachandran2018searching}, and Linear ($4C\to C$).
Following TF-MoE~\cite{hu2026tfmoe}, the F-module router pools $\mathbf{H}$ over $K$ to produce probabilities $\boldsymbol{\Pi}_{t,:}\in\mathbb{R}^{E}$ at each frame $t$. The operator $\operatorname{Top}_{n}(\boldsymbol{\Pi}_{t,:})$ selects the $n$ highest-probability experts. Their probabilities serve as weights, renormalized only when $n>1$, and are shared across all $K$ bands. Experts process the unpooled features pointwise along $C$, yielding the weighted output $\mathbf{H}_{\mathrm{out}}$.
For module $j$, the unweighted load-balancing loss, written for the F-module, is:
\begin{align}
\mathcal{L}_{\mathrm{bal}}^{(j)}
&=\mathbb{E}_{\mathrm{batch}}\!\left[
E\sum_{e=0}^{E-1}f_e^{(j)}\bar{\Pi}_e^{(j)}
\right], \label{eq:axis_balance}\\
f_e^{(j)}
&=\frac{1}{nT}\sum_{t=1}^{T}\sum_{r=1}^{n}
\mathbf{1}\!\left[e_{t,r}^{(j)}=e\right],\\
\bar{\Pi}_e^{(j)}
&=\frac{1}{T}\sum_{t=1}^{T}\Pi_{t,e}^{(j)}.
\end{align}
Here, $e_{t,r}^{(j)}$ is the $r$-th selected expert at frame $t$. Within each mixture, $f_e^{(j)}$ is expert $e$'s fraction of selected assignments~\cite{hu2026tfmoe}, $\bar{\Pi}_e^{(j)}$
is its mean routing probability over frames, and $\mathbb{E}_{\mathrm{batch}}$ denotes batch averaging.

\subsection{Sound-Event-Supervised Routing and Joint Training}
\label{ssec:event_routing}

As shown in Fig .\ref{fig:pipeline}(C.2), the event-routed FFN is the front FFN in each F- and T-module and uses a separate gate and expert bank for mixture-level routing. Pooling both axes of $\mathbf{H}$ gives:

{\allowdisplaybreaks
\begin{align}
\bar{\mathbf{h}}
&=\frac{1}{TK}\sum_{t=1}^{T}\sum_{k=1}^{K}\mathbf{H}_{t,k,:}
\in\mathbb{R}^{C}, \label{eq:event_gate}\\
\boldsymbol{\pi}
&=\operatorname{softmax}\!\left(\operatorname{Linear}_{C\to E}
 (\bar{\mathbf{h}})\right)\in\mathbb{R}^{E}.
\end{align}}%
The event router applies $\operatorname{Top}_{n}(\boldsymbol{\pi})$ to select the $n$ experts with the largest mixture-level probabilities. Their corresponding weights are shared by all $T\times K$ tokens. The weighted output has the same $T\times K\times C$ shape as the input. The gates and expert banks are independent across F/T modules and TF-MoE blocks.
 During training, a fixed mapping converts source event labels into expert IDs. Averaging the $S$ one-hot assignments yields the permutation-invariant target $\mathbf{q}\in\mathbb{R}^{E}$, where $q_e$ is the fraction of sources mapped to expert $e$. Each event gate minimizes the Kullback--Leibler divergence $D_{\mathrm{KL}}$:

\begin{equation}
\mathcal{L}_{\mathrm{evt}}^{(j)}
=\mathbb{E}_{\mathrm{batch}}\!\left[
D_{\mathrm{KL}}\!\left(
\mathbf{q}\,\middle\|\,\boldsymbol{\pi}^{(j)}
\right)\right].
\label{eq:event_supervision}
\end{equation}

This soft target preserves the source composition of a mixture while allowing the router to select $n$ experts. Inference requires only mixture features, without event labels or prompts.
The joint training objective is:
\begin{equation}
\begin{aligned}
\mathcal{L}
=\mathcal{L}_{\mathrm{sep}}
+\frac{\lambda_{\mathrm{evt}}}{2R}
\sum_{j=1}^{2R}\mathcal{L}_{\mathrm{evt}}^{(j)}+\lambda_{\mathrm{bal}}
\sum_{j=1}^{2R}\mathcal{L}_{\mathrm{bal}}^{(j)}.
\end{aligned}
\label{eq:total_objective}
\end{equation}

Here, $\mathcal{L}_{\mathrm{sep}}$ is the source- and batch-averaged permutation-invariant loss, combining waveform and STFT real/imaginary $\ell_1$ errors, negative signal-to-distortion ratio (SDR), and negative signal-to-noise ratio (SNR) with unit weights under a common source permutation. The index $j=1,\ldots,2R$ enumerates the F- and T-modules in the $R$ blocks; $\mathcal{L}_{\mathrm{evt}}^{(j)}$ and $\mathcal{L}_{\mathrm{bal}}^{(j)}$ are the event and TF load-balancing losses for their respective FFNs. We set $\lambda_{\mathrm{evt}}=1$ and $\lambda_{\mathrm{bal}}=10^{-3}$.

\section{Experiments}
\label{sec:experiment}
\subsection{Experiment Setup}
\label{ssec:expsetup}

\textbf{Dataset \& Baseline.}
Our approximately 2,000-hour training pool spans speech, singing, music, and environmental sounds. Speech comes from AISHELL-1~\cite{bu2017aishell1}, AISHELL-3~\cite{shi2021aishell3}, and LibriTTS R~\cite{koizumi2023librittsr}; singing comes from SingVERSE and VocalSet~\cite{jiang2025singverse,wilkins2018vocalset}. Other sources include AudioSet~\cite{gemmeke2017audioset}, MUSIC21~\cite{zhao2019soundmotions}, BBC Sound Effects~\cite{bbc1991soundeffects}, and VGGSound~\cite{chen2020vggsound}. To study data scaling without distribution shifts, we draw nested subsets uniformly at random: 200 hours from the full pool and 67 from the 200-hour subset. Evaluation uses the three-source Hive test partition~\cite{li2026hive}: 25,000 ten-second mixtures (75,000 target sources). All systems share 16-kHz inputs and scoring pipeline. For fair comparison, baselines include BSRNN-EDA~\cite{wangUSEUnifiedModel2026}, combining BSRNN~\cite{luo2023bsrnn} with EDA~\cite{eendeda} and trained on our dataset, pre-trained CLAPSep~\cite{ma2024clapsep} and AudioSep~\cite{liu2025audiosep}, and the \texttt{base} configuration of SAM Audio~\cite{shi2026samaudio}. Prompt-based baselines extract sources sequentially with one prompt per source.

\noindent \textbf{Training Setup.}
Training mixtures combine three distinct sources sampled online, each converted to 16~kHz mono, randomly cropped to at most 6~s, root mean square (RMS)-normalized, and independently scaled by gains uniformly sampled from $[-6.5,6.5]$ dB. In all experiments, the model is configured to separate exactly three sources, and the number of sources is fixed during inference. Shorter sources are randomly positioned through zero-padding before summation and Gaussian noise addition at an SNR uniformly sampled from $[8,20]$ dB. Event labels supervise only the auxiliary routing in event-supervised (ES) systems. The Conformer and TF-MoE models share six TF blocks, 32 channels, and 80 mel bands. TF-MoE replaces each Conformer FFN with six independent experts, supporting $n$ selections per routing unit. TF and event routing use $n=1$ (top-1) throughout, matching the corresponding dense model's active computation. ES retains this setup with mixture-level routing in the first FFN (Sec.~\ref{ssec:event_routing}). TF-MoE-L uses 48 channels, seven TF blocks, 100 bands, and 12 experts. 

Most event labels are inherited from the original annotations. For corpora without event annotations, we manually define corpus-level labels from their release descriptions and assign them using corpus identity. Each source maps to one group. After the speech-specific assignment, ordered keyword rules follow the order in Table~\ref{tab:event_groups}, merging fine labels into broad groups and resolving multiple or conflicting coarse/fine tags by the first match. Unmatched labels enter the residual group (E5 for six experts or E11 for twelve), while missing or inconsistent source-label records are rejected before training.

Averaging the source one-hot assignments preserves multiple event groups in the mixture target. Three distinct groups receive probability $1/3$ each, while repeated groups accumulate their source proportions. Top-1 limits execution to one expert per module without truncating this soft target. Independent routers can select different relevant experts across blocks.

\vspace{-9pt}
\begin{table}[!htbp]
\centering
\caption{Sound-event groups for six- and twelve-expert supervision. A dash denotes an unused ID.}
\vspace{1mm}
\label{tab:event_groups}
\begingroup
\fontsize{9}{10}\selectfont
\setlength{\tabcolsep}{3pt}
\renewcommand{\arraystretch}{1}
\begin{tabular}{@{}l>{\raggedright\arraybackslash}p{0.38\columnwidth}>{\raggedright\arraybackslash}p{0.45\columnwidth}@{}}
\toprule
\textbf{ID} & \textbf{6 experts} & \textbf{12 experts} \\
\midrule
E0 & English read speech & English read speech \\
E1 & Mandarin read speech & Mandarin news and topical reading \\
E2 & Diverse speech and vocal sounds & Neutral Mandarin utterances \\
E3 & Animal sounds & Other speech \\
E4 & Music and instruments & Singing and vocal music \\
E5 & Environmental and other sounds & Other human sounds \\
E6 & -- & Bird sounds \\
E7 & -- & Other animal sounds \\
E8 & -- & String instruments \\
E9 & -- & Other music and instruments \\
E10 & -- & Vehicles and transport \\
E11 & -- & Nature, household, and other sounds \\
\bottomrule
\end{tabular}
\endgroup
\end{table}


\noindent \textbf{Metrics.}
Signal fidelity is measured by SDR~\cite{vincent2006performance} and SI-SDR~\cite{leroux2019sdr}. SDR-based permutation-invariant matching determines a source assignment shared by all metrics. LPAPS measures deep perceptual distance (lower is better)~\cite{manor2024zeroshot}; CLAP-A and CLAP-T measure audio--audio and audio--label cosine similarity using LAION-CLAP~\cite{wu2023largescaleclap}. SAM-Audio Judge~\cite{shi2026samaudio} assesses reference-free overall quality (OQ), recall (Rec), precision (Pre), and faithfulness (Fai). Additionally, we report total parameters, GMACs for one second of 16-kHz audio, and real-time factor (RTF).

\vspace{-5pt}
\subsection{Main Results}
\label{ssec:results}
\vspace{-5pt}

\begin{table*}[t]
\centering
\caption{Separation results and computational overhead on the Hive test set (3-mix). For separation-quality metrics, \textbf{boldface} and \underline{underlining} denote
the best and second-best measured values across all rows,
respectively, with ties marked equally.
$^\dagger$Prompt-based baselines extract sources sequentially
with one prompt per source.}
\vspace{1mm}
\label{tab:main_results}
\begingroup
\renewcommand{\baselinestretch}{1}
\fontsize{9}{10.5}\selectfont
\setlength{\tabcolsep}{1pt}
\renewcommand{\arraystretch}{1}
\begin{tabular*}{\textwidth}{
@{\extracolsep{\fill}}ll*{12}{c}@{}
}
\toprule
\multirow{2}{*}{Model}
& \multirow{2}{*}{Train}
& \multicolumn{3}{c}{Computational overhead$\downarrow$}
& \multicolumn{2}{c}{Signal fidelity (dB)$\uparrow$}
& \multicolumn{3}{c}{Perceptual/semantic quality}
& \multicolumn{4}{c}{SAM-Audio Judge$\uparrow$} \\
\cmidrule(lr){3-5}
\cmidrule(lr){6-7}
\cmidrule(lr){8-10}
\cmidrule(lr){11-14}
& & \#Params. (M) & GMACs & RTF
& SDR & SI-SDR
& LPAPS$\downarrow$ & CLAP-A$\uparrow$ & CLAP-T$\uparrow$
& OQ & Rec & Pre & Fai \\
\midrule

Mixture
& - & - & - & -
& -3.50 & -3.59
& 3.98 & 0.62 & 0.21
& 1.27 & \textbf{5.00} & 1.24 & \textbf{4.83} \\

CLAPSep$^\dagger$~\cite{ma2024clapsep}
& Pretrained & 319.5 & 56.38 & 0.07
& 0.20 & -0.98
& 3.65 & 0.70 & \underline{0.23}
& 2.87 & 4.82 & 2.90 & 4.35 \\

AudioSep$^\dagger$~\cite{liu2025audiosep}
& Pretrained & 238.6 & 127.09 & 0.04
& 1.98 & 1.11
& 3.43 & 0.74 & \textbf{0.25}
& 3.03 & \underline{4.85} & 3.05 & \underline{4.41} \\

SAM-Audio$^\dagger$~\cite{shi2026samaudio}
& Pretrained & 2040.9 & 935.36 & 1.68
& -7.24 & -13.89
& 4.36 & 0.62 & 0.21
& 3.03 & 3.92 & 3.35 & 3.71 \\

BSRNN-EDA~\cite{wangUSEUnifiedModel2026}
& 2,000 h & 9.7 & 27.57 & 0.02
& \underline{8.22} & \underline{5.68}
& \textbf{2.98} & \textbf{0.82} & \underline{0.23}
& \textbf{3.57} & 4.72 & \textbf{3.66} & 4.30 \\
\midrule

Conformer
& 67 h & 4.9 & 1.35 & 0.05
& 6.52 & 3.73
& 3.45 & 0.74 & 0.22
& 3.09 & 4.81 & 3.12 & 4.30 \\

Conformer
& 200 h & 4.9 & 1.35 & 0.05
& 7.19 & 4.39
& 3.34 & 0.76 & 0.22
& 3.24 & 4.79 & 3.29 & 4.29 \\

Conformer-L
& 200 h & 8.9 & 4.18 & 0.06
& 7.54 & 4.76
& 3.30 & 0.77 & 0.22
& 3.27 & 4.80 & 3.31 & 4.32 \\

TF-MoE
& 67 h & 5.9 & 1.35 & 0.07
& 6.77 & 4.02
& 3.41 & 0.75 & 0.22
& 3.14 & 4.81 & 3.17 & 4.29 \\

\hspace{1.5em}+ ES
& 67 h & 5.9 & 1.35 & 0.07
& 6.85 & 4.08
& 3.41 & 0.75 & 0.22
& 3.17 & 4.80 & 3.20 & 4.29 \\

TF-MoE
& 200 h & 5.9 & 1.35 & 0.07
& 7.50 & 4.77
& 3.30 & 0.77 & 0.22
& 3.29 & 4.78 & 3.35 & 4.27 \\

\hspace{1.5em}+ ES
& 200 h & 5.9 & 1.35 & 0.07
& 7.53 & 4.80
& 3.27 & 0.77 & 0.22
& 3.32 & 4.77 & 3.37 & 4.28 \\

TF-MoE-L
& 200 h & 14.7 & 4.18 & 0.09
& 7.82 & 5.14
& 3.23 & 0.78 & 0.22
& 3.36 & 4.78 & 3.42 & 4.30 \\

\hspace{1.5em}+ ES
& 200 h & 14.7 & 4.18 & 0.09
& 7.94 & 5.26
& 3.21 & 0.79 & 0.22
& 3.37 & 4.77 & 3.43 & 4.29 \\

\textbf{TF-MoE-L}
& 2,000 h & 14.7 & 4.18 & 0.09
& 8.04 & 5.42
& 3.19 & 0.79 & \underline{0.23}
& 3.38 & 4.78 & 3.44 & 4.30 \\

\hspace{1.5em}\textbf{+ ES}
& 2,000 h & 14.7 & 4.18 & 0.09
& \textbf{8.42} & \textbf{5.76}
& \underline{3.12} & \underline{0.80} & \underline{0.23}
& \underline{3.44} & 4.76 & \underline{3.50} & 4.29 \\

\bottomrule
\end{tabular*}
\endgroup
\vspace{-10pt}
\end{table*}

\noindent\textbf{Regular TF-MoE.}
Table~\ref{tab:main_results} shows that TF-MoE offers a favorable capacity--computation trade-off for universal sound separation. TF-MoE-L requires substantially less computation than BSRNN-EDA, reducing GMACs by 84.8\%. Compared with prompt-based baselines, it also uses fewer parameters and GMACs, although differences in pretraining data and objectives limit direct comparability. However, arithmetic savings do not imply faster execution: TF-MoE-L has a higher RTF than BSRNN-EDA, with the current implementation contributing to its latency. 

\begin{figure}[!htp]
    \centering
    \includegraphics[width=\columnwidth]{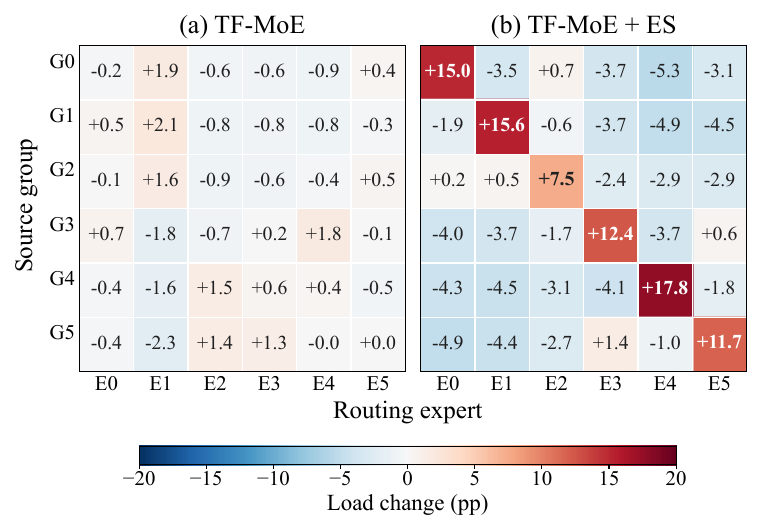}
    \vspace{-20pt}
    \caption{Class-conditioned shifts in expert load relative to global means on 2,048 three-source mixtures: (a) without and (b) with event supervision. Loads are averaged across the F-module FFN1 routers in all six blocks. Class groups $G0$--$G5$ follow Table \ref{tab:event_groups}.}
    \vspace{-20pt}
    \label{fig:class_routing}
\end{figure}

\noindent\textbf{Event supervising.}
ES consistently improves SDR, SI-SDR, OQ, and precision across evaluated data and model scales, without inference-time labels or additional reported inference cost. For TF-MoE-L, SDR and SI-SDR gains increase with training data. With 12 event experts and ES, TF-MoE-L trained on 2,000~h outperforms BSRNN-EDA trained on the same data duration in separation metrics.
Fig.~\ref{fig:class_routing} compares the expert-selection patterns learned with and without event supervision. The two $6\times6$ heatmaps use source-event groups as rows and routing experts as columns. Each cell reports the group-conditioned expert load minus the global expert load in percentage points. The Hive test partition lacks some training-defined event groups (i.e., Mandarin read speech) , so the figure uses a held-out set constructed with the six groups and balanced three-source mixtures. With load-balancing loss alone, the load shifts across the heatmap are generally close to zero, indicating that the routing pattern is only weakly related to the source-event group. With event supervision, the diagonal entries become clearly positive, with increases of approximately 7--18 percentage points for the corresponding experts. Therefore, event supervision makes the router prefer event-matched expert networks while keeping top-1 routing and the associated inference computation unchanged. Recorded routing examples further show that experts for all constituent event groups can be activated across different blocks, despite selecting only one expert per module. The demo visualizes routing probabilities and expert selections across blocks.


\vspace{-10pt}
\subsection{Data \& Model Scaling}
\label{ssec:scaling}
\vspace{-5pt}

\noindent\textbf{Data scaling.}
Nested subsets preserve the mixture construction and evaluation
protocol.
Quality improves from 67 to 200~h; further scaling TF-MoE-L
to 2,000~h yields 0.22/0.28~dB SDR/SI-SDR gains without ES
and 0.48/0.50~dB with ES.
Although ES gains are not strictly monotonic across data sizes,
the stronger benefit at 2,000~h suggests sufficient training
coverage helps event specialization generalize reliably,
consistent with more robust separation across diverse test mixtures.

\noindent\textbf{Model scaling.}
At 200~h, expanding TF-MoE to TF-MoE-L improves SDR/SI-SDR
by 0.32/0.37~dB without ES and 0.41/0.46~dB with ES.
More channels, depth, frequency bands, and experts increase
GMACs from 1.35 to 4.18 while retaining top-1 routing.
With ES, data expansion to 2,000~h yields slightly larger
gains than model scaling at 200~h without increasing inference
cost, highlighting complementary benefits from capacity and data.
\vspace{-10pt}

\section{Conclusion}
\label{sec:conclusion}
\vspace{-5pt}
TF-MoE-USS combines axis-level and event-supervised mixture-level routing to improve separation at approximately matched MAC counts within each model configuration. 
Experiments demonstrate improved separation through increased conditional capacity without proportional computational growth, while event supervision promotes specialization without inference-time labels. 
Future work will investigate routing robustness to imperfect annotations and more efficient sparse implementations to translate these gains into practical deployment benefits.

\bibliographystyle{IEEEbib}
\bibliography{refs}

@inproceedings{wang2023tfgridnet,
  author = {Wang, Zhong-Qiu and others},
  title = {{TF-GridNet}: Making Time-Frequency Domain Models Great Again for Monaural Speaker Separation},
  booktitle = {ICASSP},
  year = {2023},
  pages = {1--5},
  doi = {10.1109/icassp49357.2023.10094992},
  url = {https://doi.org/10.1109/icassp49357.2023.10094992},
}

@inproceedings{zhao2024amaeuss,
  author = {Zhao, Junqi and Liu, Xubo and Zhao, Jinzheng and Yuan, Yi and Kong, Qiuqiang and Plumbley, Mark D. and Wang, Wenwu},
  title = {Universal Sound Separation with Self-Supervised Audio Masked Autoencoder},
  booktitle = {EUSIPCO},
  year = {2024},
  pages = {1--5},
  doi = {10.23919/EUSIPCO63174.2024.10715391},
  url = {https://eurasip.org/Proceedings/Eusipco/Eusipco2024/pdfs/0000001.pdf}
}

@article{luo2023bsrnn,
  author = {Luo, Yi and others},
  title = {Music Source Separation With Band-Split {RNN}},
  journal = {IEEE/ACM TASLP},
  year = {2023},
  pages = {1893--1901},
  doi = {10.1109/taslp.2023.3271145},
  url = {https://doi.org/10.1109/taslp.2023.3271145},
  volume = {31},
}

@inproceedings{kavalerov2019universal,
  author = {Kavalerov, Ilya and others},
  title = {Universal Sound Separation},
  booktitle = {WASPAA},
  year = {2019},
  pages = {175--179},
  doi = {10.1109/waspaa.2019.8937253},
  url = {https://doi.org/10.1109/waspaa.2019.8937253},
}

@inproceedings{wisdom2021fuss,
  author = {Wisdom, Scott and others},
  title = {What's All the {FUSS} About Free Universal Sound Separation Data?},
  booktitle = {ICASSP},
  year = {2021},
  pages = {186--190},
  doi = {10.1109/icassp39728.2021.9414774},
  url = {https://doi.org/10.1109/icassp39728.2021.9414774},
}

@inproceedings{wisdom2020mixit,
  author = {Scott Wisdom and others},
  title = {Unsupervised Sound Separation Using Mixture Invariant Training},
  booktitle = {NeurIPS},
  volume = {33},
  pages = {3846--3857},
  year = {2020},
  url = {https://papers.nips.cc/paper/2020/hash/28538c394c36e4d5ea8ff5ad60562a93-Abstract.html},
}

@inproceedings{li2026hive,
  author = {Kai Li and others},
  title = {A Semantically Consistent Dataset for Data-Efficient Query-Based Universal Sound Separation},
  booktitle = {ICML},
  year = {2026},
  url = {https://icml.cc/virtual/2026/poster/60936},
}

@article{liu2025audiosep,
  author = {Liu, Xubo and others},
  title = {Separate Anything You Describe},
  journal = {IEEE TASLPRO},
  year = {2025},
  pages = {458--471},
  doi = {10.1109/taslp.2024.3520017},
  url = {https://doi.org/10.1109/taslp.2024.3520017},
  volume = {33},
}

@inproceedings{perez2018film,
  author    = {Perez, Ethan and others},
  title     = {{FiLM}: Visual Reasoning with a General Conditioning Layer},
  booktitle = {AAAI},
  volume    = {32},
  number    = {1},
  pages     = {3942--3951},
  year      = {2018},
  doi       = {10.1609/aaai.v32i1.11671}
}

@article{ma2024clapsep,
  author = {Ma, Hao and others},
  title = {{CLAPSep}: Leveraging Contrastive Pre-Trained Model for Multi-Modal Query-Conditioned Target Sound Extraction},
  journal = {IEEE/ACM TASLP},
  year = {2024},
  pages = {4945--4960},
  doi = {10.1109/taslp.2024.3497586},
  url = {https://doi.org/10.1109/taslp.2024.3497586},
  volume = {32},
}

@inproceedings{yuan2025flowsep,
  author = {Yuan, Yi and others},
  title = {{FlowSep}: Language-Queried Sound Separation with Rectified Flow Matching},
  booktitle = {ICASSP},
  year = {2025},
  pages = {1--5},
  doi = {10.1109/icassp49660.2025.10890129},
  url = {https://doi.org/10.1109/icassp49660.2025.10890129},
}

@inproceedings{shi2026samaudio,
  author = {Bowen Shi and others},
  title = {{SAM Audio}: Segment Anything in Audio},
  booktitle = {ICML},
  year = {2026},
  url = {https://icml.cc/virtual/2026/poster/64195},
}

@inproceedings{shazeer2017moe,
  author = {Noam Shazeer and others},
  title = {Outrageously Large Neural Networks: The Sparsely-Gated Mixture-of-Experts Layer},
  booktitle = {ICLR},
  year = {2017},
  url = {https://openreview.net/forum?id=B1ckMDqlg},
}

@article{fedus2022switch,
  author = {William Fedus and others},
  title = {Switch Transformers: Scaling to Trillion Parameter Models with Simple and Efficient Sparsity},
  journal = {JMLR},
  volume = {23},
  number = {120},
  pages = {1--39},
  year = {2022},
  url = {https://jmlr.org/papers/v23/21-0998.html},
}

@inproceedings{hu2026tfmoe,
  author = {Qinzhe Hu and others},
  title = {{TF-MoE}: Time-Frequency Mixture-of-Experts for Efficient Speech Separation},
  booktitle = {Interspeech},
  year = {2026},
  url = {https://www.microsoft.com/en-us/research/publication/tf-moe-time-frequency-mixture-of-experts-for-efficient-speech-separation/},
}

@inproceedings{tzinis2020classification,
  author = {Tzinis, Efthymios and others},
  title = {Improving Universal Sound Separation Using Sound Classification},
  booktitle = {ICASSP},
  year = {2020},
  pages = {96--100},
  doi = {10.1109/icassp40776.2020.9053921},
  url = {https://doi.org/10.1109/icassp40776.2020.9053921},
}

@inproceedings{gulati2020conformer,
  author = {Anmol Gulati and others},
  title = {{Conformer}: Convolution-augmented Transformer for Speech Recognition},
  booktitle = {Interspeech},
  year = {2020},
  pages = {5036--5040},
  doi = {10.21437/Interspeech.2020-3015},
  url = {https://www.isca-archive.org/interspeech_2020/gulati20_interspeech.html}
}

@inproceedings{ramachandran2018searching,
  author    = {Prajit Ramachandran and others},
  title     = {Searching for Activation Functions},
  booktitle = {ICLR Workshops},
  year      = {2018},
  url       = {https://openreview.net/forum?id=Hkuq2EkPf}
}

@inproceedings{bu2017aishell1,
  author = {Bu, Hui and others},
  title = {{AISHELL-1}: An Open-Source Mandarin Speech Corpus and a Speech Recognition Baseline},
  booktitle = {O-COCOSDA},
  year = {2017},
  pages = {1--5},
  doi = {10.1109/ICSDA.2017.8384449}
}

@inproceedings{shi2021aishell3,
  author = {Shi, Yao and others},
  title = {{AISHELL-3}: A Multi-Speaker Mandarin {TTS} Corpus},
  booktitle = {Interspeech},
  year = {2021},
  pages = {2756--2760},
  doi = {10.21437/Interspeech.2021-755}
}

@inproceedings{gemmeke2017audioset,
  author = {Gemmeke, Jort F. and others},
  title = {Audio Set: An Ontology and Human-Labeled Dataset for Audio Events},
  booktitle = {ICASSP},
  year = {2017},
  pages = {776--780},
  doi = {10.1109/ICASSP.2017.7952261}
}

@inproceedings{koizumi2023librittsr,
  author = {Koizumi, Yuma and others},
  title = {{LibriTTS-R}: A Restored Multi-Speaker Text-to-Speech Corpus},
  booktitle = {Interspeech},
  year = {2023},
  pages = {5496--5500},
  doi = {10.21437/Interspeech.2023-1584}
}

@inproceedings{zhao2019soundmotions,
  author = {Zhao, Hang and others},
  title = {The Sound of Motions},
  booktitle = {ICCV},
  year = {2019},
  pages = {1735--1744},
  doi = {10.1109/ICCV.2019.00182}
}

@article{jiang2025singverse,
  author = {Jiang, Shaohan and others},
  title = {{SingVERSE}: A Diverse, Real-World Benchmark for Singing Voice Enhancement},
  journal = {arXiv preprint arXiv:2509.20969},
  year = {2025},
  url = {https://arxiv.org/abs/2509.20969}
}

@misc{bbc1991soundeffects,
  author = {{British Broadcasting Corporation}},
  title = {{BBC Sound Effects Library}},
  howpublished = {Films for the Humanities \& Sciences},
  year = {1991}
}

@inproceedings{chen2020vggsound,
  author = {Chen, Honglie and others},
  title = {{VGGSound}: A Large-Scale Audio-Visual Dataset},
  booktitle = {ICASSP},
  year = {2020},
  pages = {721--725},
  doi = {10.1109/ICASSP40776.2020.9053174}
}

@inproceedings{wilkins2018vocalset,
  title={{VocalSet}: A Singing Voice Dataset},
  author={Wilkins, Julia and Seetharaman, Prem and Wahl, Alison and Pardo, Bryan},
  booktitle={ISMIR},
  pages={468--474},
  year={2018},
}

@article{vincent2006performance,
  author = {Vincent, Emmanuel and others},
  title = {Performance Measurement in Blind Audio Source Separation},
  journal = {IEEE TASLP},
  year = {2006},
  volume = {14},
  number = {4},
  pages = {1462--1469},
  doi = {10.1109/TSA.2005.858005}
}

@inproceedings{leroux2019sdr,
  author = {{Le Roux}, Jonathan and others},
  title = {{SDR} -- Half-Baked or Well Done?},
  booktitle = {ICASSP},
  year = {2019},
  pages = {626--630},
  doi = {10.1109/ICASSP.2019.8683855}
}

@inproceedings{manor2024zeroshot,
  author = {Manor, Hila and others},
  title = {Zero-Shot Unsupervised and Text-Based Audio Editing Using {DDPM} Inversion},
  booktitle = {ICML},
  year = {2024},
  volume = {235},
  pages = {34603--34629},
  url = {https://proceedings.mlr.press/v235/manor24a.html}
}

@inproceedings{wu2023largescaleclap,
  author = {Wu, Yusong and others},
  title = {Large-Scale Contrastive Language-Audio Pretraining with Feature Fusion and Keyword-to-Caption Augmentation},
  booktitle = {ICASSP},
  year = {2023},
  pages = {1--5},
  doi = {10.1109/ICASSP49357.2023.10095969}
}

@article{eendeda,
  title={Encoder-decoder based attractors for end-to-end neural diarization},
  author={Horiguchi, Shota and others},
  journal={IEEE/ACM TASLP},
  volume={30},
  pages={1493--1507},
  year={2022},
  publisher={IEEE}
}

@article{wangUSEUnifiedModel2026,
  title = {{{USE}}: {{A Unified Model}} for {{Universal Sound Separation}} and {{Extraction}}},
  shorttitle = {{{USE}}},
  author = {Wang, Hongyu and others},
  year = 2026,
  month = mar,
  journal = {AAAI},
  volume = {40},
  number = {39},
  pages = {33476--33484},
  issn = {2374-3468},
  doi = {10.1609/aaai.v40i39.40635},
  urldate = {2026-04-14},
  copyright = {Copyright (c) 2026 Association for the Advancement of Artificial Intelligence},
  langid = {english},
}

@inproceedings{wisdom2021sparsemixit,
  author    = {Wisdom, Scott and Jansen, Aren and Weiss, Ron J.
               and Erdogan, Hakan and Hershey, John R.},
  title     = {Sparse, Efficient, and Semantic Mixture Invariant
               Training: Taming In-the-Wild Unsupervised Sound Separation},
  booktitle = {WASPAA},
  year      = {2021},
  pages     = {51--55},
  doi       = {10.1109/WASPAA52581.2021.9632714}
}

@inproceedings{lee2025deftmamba,
  author    = {Lee, Dongheon and Choi, Jung-Woo},
  title     = {{DeFT-Mamba}: Universal Multichannel Sound Separation
               and Polyphonic Audio Classification},
  booktitle = {ICASSP},
  year      = {2025},
  pages     = {1--5},
  doi       = {10.1109/ICASSP49660.2025.10890324}
}

@inproceedings{High-Yu2023,
  title={High Fidelity Speech Enhancement with Band-split {RNN}},
  author={Yu, Jianwei and Chen, Hangting and Luo, Yi and Gu, Rongzhi and Weng, Chao},
  booktitle={Interspeech},
  pages={2483--2487},
  year={2023},
}

@inproceedings{pons2024gass,
  author = {Pons, Jordi and Liu, Xiaoyu and Pascual, Santiago and Serr{\`a}, Joan},
  title = {{GASS}: Generalizing Audio Source Separation with Large-Scale Data},
  booktitle = {ICASSP},
  year = {2024},
  pages = {546--550},
  doi = {10.1109/ICASSP48485.2024.10446601},
  url = {https://doi.org/10.1109/ICASSP48485.2024.10446601}
}

@article{kong2026universalweak,
  author = {Kong, Qiuqiang and Chen, Ke and Liu, Haohe and Du, Xingjian and Berg-Kirkpatrick, Taylor and Dubnov, Shlomo and Plumbley, Mark D.},
  title = {Universal Source Separation With Weakly Labelled Data for Computational Auditory Scene Analysis},
  journal = {IEEE TASLPRO},
  year = {2026},
  volume = {34},
  pages = {206--219},
  doi = {10.1109/TASLPRO.2025.3636294},
  url = {https://doi.org/10.1109/TASLPRO.2025.3636294}
}

@inproceedings{munakata2021multiple,
  author = {Munakata, Hokuto and Takeda, Ryu and Komatani, Kazunori},
  title = {Multiple-Embedding Separation Networks: Sound Class-Specific Feature Extraction for Universal Sound Separation},
  booktitle = {APSIPA ASC},
  year = {2021},
  pages = {961--967},
  url = {https://www.apsipa.org/proceedings/2021/pdfs/0000961.pdf}
}

@inproceedings{paissan2025fastuss,
  author = {Paissan, Francesco and Wichern, Gordon and Masuyama, Yoshiki and Aihara, Ryo and Germain, Fran{\c{c}}ois G. and Saijo, Kohei and {Le Roux}, Jonathan},
  title = {{FasTUSS}: Faster Task-Aware Unified Source Separation},
  booktitle = {WASPAA},
  year = {2025},
  pages = {1--5},
  doi = {10.1109/WASPAA66052.2025.11230943},
  url = {https://www.merl.com/publications/TR2025-143}
}

@inproceedings{saijo2025tuss,
  author = {Saijo, Kohei and Ebbers, Janek and Germain, Fran{\c{c}}ois G. and Wichern, Gordon and {Le Roux}, Jonathan},
  title = {Task-Aware Unified Source Separation},
  booktitle = {ICASSP},
  year = {2025},
  pages = {1--5},
  doi = {10.1109/ICASSP49660.2025.10887819},
  url = {https://www.merl.com/publications/TR2025-032}
}

@inproceedings{postolache2023adversarial,
  author = {Postolache, Emilian and Pons, Jordi and Pascual, Santiago and Serr{\`a}, Joan},
  title = {Adversarial Permutation Invariant Training for Universal Sound Separation},
  booktitle = {ICASSP},
  year = {2023},
  pages = {1--5},
  doi = {10.1109/ICASSP49357.2023.10096847},
  url = {https://doi.org/10.1109/ICASSP49357.2023.10096847}
}

@inproceedings{li2023crossmodality,
  author = {Li, Chenda and Qian, Yao and Chen, Zhuo and Wang, Dongmei and Yoshioka, Takuya and Liu, Shujie and Qian, Yanmin and Zeng, Michael},
  title = {Target Sound Extraction with Variable Cross-Modality Clues},
  booktitle = {ICASSP},
  year = {2023},
  pages = {1--5},
  doi = {10.1109/ICASSP49357.2023.10095266},
  url = {https://doi.org/10.1109/ICASSP49357.2023.10095266}
}

@inproceedings{kong2018jointseparation,
  author = {Kong, Qiuqiang and Xu, Yong and Wang, Wenwu and Plumbley, Mark D.},
  title = {A Joint Separation-Classification Model for Sound Event Detection of Weakly Labelled Data},
  booktitle = {ICASSP},
  year = {2018},
  pages = {321--325},
  doi = {10.1109/ICASSP.2018.8462448},
  url = {https://doi.org/10.1109/ICASSP.2018.8462448}
}

@inproceedings{chen2022zeroshotuss,
  author = {Chen, Ke and Du, Xingjian and Zhu, Bilei and Ma, Zejun and Berg-Kirkpatrick, Taylor and Dubnov, Shlomo},
  title = {Zero-Shot Audio Source Separation through Query-Based Learning from Weakly-Labeled Data},
  booktitle = {AAAI},
  year = {2022},
  volume = {36},
  pages = {4441--4449},
  doi = {10.1609/aaai.v36i4.20366},
  url = {https://ojs.aaai.org/index.php/AAAI/article/view/20366}
}

\end{document}